\documentclass[twocolumn,aps,showpacs,amssymb,floatfix]{revtex4}
\newcommand{\be}{\begin{equation}}
\newcommand{\ee}{\end{equation}}
\newcommand{\beq}{\begin{eqnarray}}
\newcommand{\eeq}{\end{eqnarray}}
\usepackage{bm}
\usepackage{graphicx}%
\usepackage{epsf}
\usepackage{epsfig}

\usepackage{amsmath}
\usepackage{amsfonts,amssymb}

\begin{document}

\title{Axial Nucleon to $\Delta$ transition form factors on 2+1 flavor hybrid lattices}
\author{C.~Alexandrou~$^a$, G. Koutsou,$^a$, Th. Leontiou,$^a$ J. W. Negele~$^b$, and A. Tsapalis~$^{c,d}$}
\affiliation{{$^a$ Department of Physics, University of Cyprus, CY-1678 Nicosia, Cyprus}\\
{$^b$
Center for Theoretical Physics, Laboratory for
 Nuclear Science and Department of Physics, Massachusetts Institute of
Technology, Cambridge, Massachusetts 02139, U.S.A.}\\{$^c$
Institute of Accelerating Systems and Applications, University of Athens,
Athens, Greece}\\
{$^d$ Hellenic Naval Academy, Hatzikyriakou Ave, Pireaus, GR 18539, Greece}
}

\date{\today}%

\newcommand{\twopt}[5]{\langle G_{#1}^{#2}(#3;\mathbf{#4};\Gamma_{#5})\rangle}
\newcommand{\threept}[7]{\langle G_{#1}^{#2}(#3,#4;\mathbf{#5},\mathbf{#6};\Gamma_{#7})\rangle}
\newcommand{\GPNN}{$G_{\pi NN}(Q^2)$}
\newcommand{\GPND}{$G_{\pi N\Delta}(Q^2)$}
\newcommand{\GA}{$G_A(Q^2)$}
\newcommand{\GP}{$G_p(Q^2)$}

 \begin{abstract}
 
We correct the values of the dominant Nucleon to $\Delta(1232)$ axial 
transition form factors $C_5^{A}(Q^2)$ and $C_6^{A}(Q^2)$ published in Ref.
C. Alexandrou {\it et al.}, Phys. Rev. D {\bf 76}, 094511
(2007). The analysis error affects only the values obtained when using
 the hybrid
action. The error affects mainly results in the low-$Q^2$ regime bringing
them into agreement with those obtained with Wilson fermions.

\end{abstract}
\maketitle

In Ref.~\cite{paper}  
results on the nucleon and nucleon to $\Delta$ 
axial and pseudoscalar  form factors within lattice QCD were presented. 
The form factors
were evaluated in the  quenched theory, using two degenerate flavors of Wilson 
fermions and within a hybrid approach that used  domain wall valence quarks
and  two degenerate flavors of
light and a strange staggered sea quarks simulated 
with the Asqtad improved action by
the MILC collaboration.
A general consistency of the calculated $Q^2$ dependence 
of the form factors 
was observed within the different discretization schemes for similar 
pion masses except for the case of the $C_5^A$ transition form factor 
at the low $Q^2$ ($ < 0.5~{\rm GeV}^2$) regime
(see Fig.17 in Ref.~\cite{paper}).
This behavior resulted from an
analysis error that affects mainly $C_5^A$ in the low $Q^2$ regime and
to a much less degree $C_6^A$.
The fact that this error affects mainly the low-$Q^2$ values for   $C_5^A$ is
understood from the structure of the equations and is due to the presence of a
denominator proportional to the transfer momentum
vector.

\begin{figure}[ht]
\epsfxsize=9.3truecm \epsfysize=11truecm
\mbox{\epsfbox{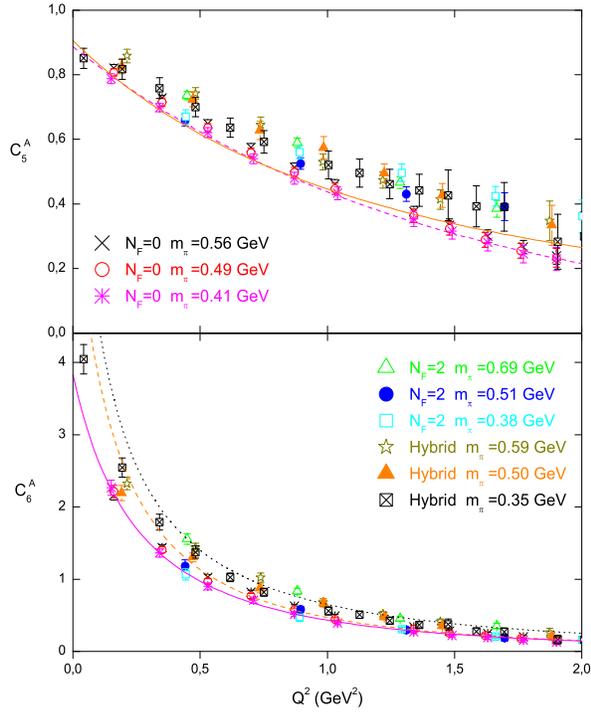}}
\caption{The upper graph shows $C_5^A(Q^2)$ and the lower graph
 $C_6^A(Q^2)$ as a function of $Q^2$.
 The notation and curves are identical to those of Fig.~17 in Ref.~\cite{paper}.
}
\label{fig:c5}
\end{figure}

\begin{figure}[ht]
\epsfxsize=9.3truecm \epsfysize=7truecm
\mbox{\epsfbox{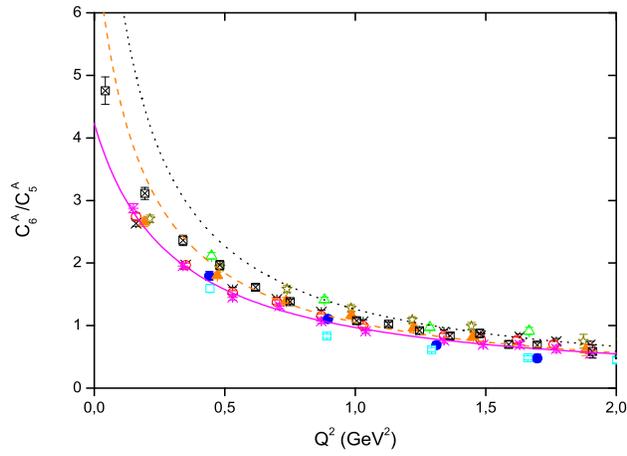}}
\caption{The ratio of $N$ to $\Delta$ axial transition 
 form factors $C_6^A(Q^2)/C_5^A(Q^2)$. 
The notation and curves are identical to those of Fig.~15 in Ref.~\cite{paper}.
}
\label{fig:c6c5}
\end{figure}

We provide in Table I the corrected data for $C_5^A/Z_A$ and $C_6^A/Z_A$
which should replace  Table IX in Ref.~\cite{paper}. We show 
in Fig.~{\ref{fig:c5},} which replaces Fig.~17 in Ref.~\cite{paper}
 the corrected data for $C_5^A$ and $C_6^A$. As can be seen the dependence
on $Q^2$ is now consistent among the different discretized actions. 
We point out that for $C_6^A$ only the values at two-$Q^2$ values
 have moved  outside error bars
 while results for $C_5^A$  obtained in the hybrid scheme have changed in the
region
$Q^2 < 0.5$ GeV$^2$. In Fig.~{\ref{fig:c6c5}}, which replaces Fig.~15 
of Ref.~\cite{paper} we provide the corrected ratio
$C_6^A / C_5^A$ and  comparison to the pion pole dominance prediction.
The correction on the values of $C_5^A$  affects also the 
Goldberger-Treiman relations studied in Ref.~\cite{paper} in the 
$Q^2 < 0.5$ GeV$^2$ region for the case of the hybrid action.
Since the general qualitative conclusions are not affected, 
we refer to Ref.~\cite{lat09} for the
corrected figures.
The rest of the conclusions drawn in Ref.~\cite{paper} remain valid.

\begin{center}
\begin{table}[ht]
\begin{tabular}{ccc}
 \hline \multicolumn{3}{c}
{N to $\Delta$: Hybrid action} \\
\hline 
$ Q^2$~(GeV$^2$)& $ C_5^A/Z_A $ & $ C_6^A/Z_A $ \\
\hline \multicolumn{3}{c} {  $m_\pi=0.594(1)$~(GeV)} \\
\hline
  0.213  &  0.774(19)   &  2.096(77)  \\
  0.482  &  0.669(16)   &  1.297(54)  \\
  0.738  &  0.582(22)   &  0.921(58)  \\
  0.983  &  0.478(22)   &  0.610(47)  \\
  1.218  &  0.427(22)   &  0.467(32)  \\
  1.445  &  0.373(27)   &  0.370(34)  \\
  1.874  &  0.313(55)   &  0.235(56)  \\
  2.079  &  0.254(46)   &  0.186(42)  \\
  2.278  &  0.175(45)   &  0.116(40)  \\
  2.472  &  0.166(55)   &  0.106(41)  \\
  2.660  &  0.173(107)  &  0.142(93)  \\
\hline \multicolumn{3}{c} { $m_\pi=0.498(3)$~(GeV)} \\
\hline
  0.191 & 0.752(21)  &    1.996(96)  \\
  0.471 & 0.658(21)  &    1.186(56)  \\
  0.735 & 0.571(27)  &    0.794(57)  \\
  0.985 & 0.522(31)  &    0.609(52)  \\
  1.224 & 0.450(27)  &    0.434(35)  \\
  1.452 & 0.387(34)  &    0.321(35)  \\
  1.882 & 0.304(56)  &    0.186(43)  \\
  2.087 & 0.298(76)  &    0.179(51)  \\
  2.284 & 0.252(148) &    0.133(91)  \\
  2.476 & 0.182(170) &    0.101(112) \\
\hline \multicolumn{3}{c} { $m_\pi=0.357(2)$~(GeV)} \\
\hline
  0.042(16) & 0.785(29) &   3.730(188)  \\
  0.194(14) & 0.753(29) &   2.346(122)  \\
  0.341(8)  & 0.699(30) &   1.651(102)  \\
  0.482(9)  & 0.645(27) &   1.268(80 )  \\
  0.619(8)  & 0.587(27) &   0.946(54)   \\
  0.751(9)  & 0.546(32) &   0.754(50)   \\
  1.005(11) & 0.480(40) &   0.518(53)   \\
  1.127(16) & 0.457(40) &   0.467(45)   \\
  1.246(16) & 0.425(42) &   0.391(43)   \\
  1.362(23) & 0.407(46) &   0.340(42)   \\
  1.475(51) & 0.393(72) &   0.344(70)   \\
  1.586(18) & 0.361(56) &   0.253(47)   \\
  1.695(35) & 0.360(69) &   0.248(50)   \\
  1.906(65) & 0.261(78) &   0.153(56)   \\

\hline
\end{tabular}
\label{Table:Hybrid results NDelta}
\caption{The first column gives the $Q^2$ in GeV$^2$, the second $C_5^A/Z_A$ 
and the third $C_6^A/Z_A$. 
The errors quoted are jackknife errors. We provide also the error on $Q^2$
for the large lattice due to the $m_N$ and $m_{\Delta}$ uncertainty.}
\end{table}
\end{center}

\end{document}